%% file: Understanding_the_Use_of_a_Large_Language_Model-Powered_Guide_to_Make_Virtual_Reality_Accessible_for_Blind_and_Low_Vision_People.tex
\documentclass[sigconf]{acmart}

\AtBeginDocument{%
  }

\usepackage{geometry}
\usepackage{booktabs}
\usepackage{multirow}
\PassOptionsToPackage{table,xcdraw}{xcolor}
\usepackage{xcolor}
\copyrightyear{2026}
\acmYear{2026}
\setcopyright{cc}
\setcctype{by-nc-nd}
\acmConference[CHI '26]{Proceedings of the 2026 CHI Conference on Human Factors in Computing Systems}{April 13--17, 2026}{Barcelona, Spain}
\acmBooktitle{Proceedings of the 2026 CHI Conference on Human Factors in Computing Systems (CHI '26), April 13--17, 2026, Barcelona, Spain}
\acmPrice{}
\acmDOI{10.1145/3772318.3791143}
\acmISBN{979-8-4007-2278-3/2026/04}

\usepackage{subcaption}
\usepackage{array,ragged2e}
\newcolumntype{P}[1]{>{\RaggedRight\arraybackslash}p{#1}}
\newcolumntype{M}[1]{>{\centering\arraybackslash}m{#1}}

\begin{document}


\title{Understanding the Use of a Large Language Model-Powered Guide to Make Virtual Reality Accessible for Blind and Low Vision People}

\author{Jazmin Collins}
\email{jc2884@cornell.edu}
\affiliation{%
  \institution{Cornell University}
  \city{New York}
  \state{New York}
  \country{USA}}

%<TODO> Revise your own affiliation

%<TODO> Revise your own affiliation

\author{Sharon Lin}
\email{syl55@cornell.edu}
\affiliation{%
  \institution{Cornell Tech}
  \city{New York}
  \state{New York}
  \country{USA}
}

\author{Tianqi Liu}
\email{tl889@cornell.edu}
\affiliation{%
  \institution{Cornell Tech}
  \city{New York}
  \state{New York}
  \country{USA}
}

\author{Andrea Stevenson Won}
\email{asw248@cornell.edu}
\affiliation{%
  \institution{Cornell University}
  \city{Ithaca}
  \state{New York}
  \country{USA}
}

\author{Shiri Azenkot}
\email{shiri.azenkot@cornell.edu}
\affiliation{%
  \institution{Cornell Tech}
  \city{New York}
  \state{New York}
  \country{USA}
}

\renewcommand{\shortauthors}{Collins et al.}

\input{1Abstract}



\begin{teaserfigure}
 \includegraphics[width=\linewidth]{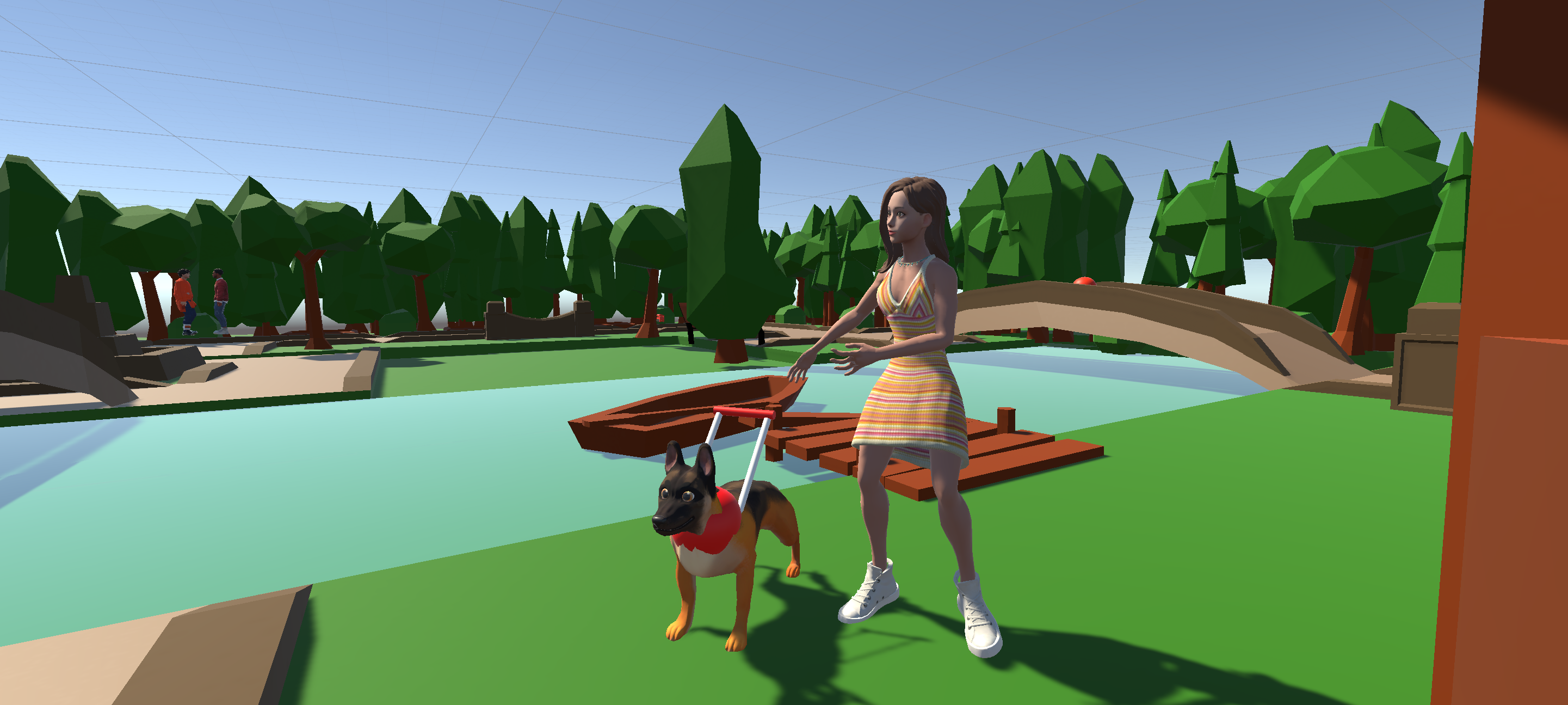}
  \caption{A screenshot of a participant utilizing the AI guide in its dog form, traveling through a virtual park together.}
  \label{fig:teaser}
  \Description{A screenshot of a virtual park environment. A female avatar and a German Shepherd breed dog with a bright red guide dog harness stand in the center of the image. A wooden dock is behind them at the edge of a river cutting through the park. A bridge crosses the river to the woman's right. In the distance, there is a line of green trees.}
\end{teaserfigure}

\maketitle

\input{2Introduction}
\input{3RelatedWork}
\input{4Methods}
\input{5Findings}
\input{6Discussion}
\input{7Conclusion}

\begin{acks}
  This paper is based upon work supported in part by the National Science Foundation under Grant No. 1942693. The contents of this paper do not necessarily represent the policy of the funders, and no endorsement should be assumed. We thank all the participants for their time.
  We would also like to thank the LightHouse for the Blind and Visually Impaired for their efforts in testing in-progress prototypes, spearheading recruitment, and proctoring our study sessions.
\end{acks}

\bibliographystyle{ACM-Reference-Format}
\bibliography{references}

%%\appendix

\end{document}

%% file: 1Abstract.tex
\begin{abstract}
    As social virtual reality (VR) grows more popular, addressing accessibility for blind and low vision (BLV) users is increasingly critical. Researchers have proposed an AI “sighted guide” to help users navigate VR and answer their questions, but it has not been studied with users. To address this gap, we developed a large language model (LLM)-powered guide and studied its use with 16 BLV participants in virtual environments with confederates posing as other users. We found that when alone, participants treated the guide as a tool, but treated it companionably around others, giving it nicknames, rationalizing its mistakes with its appearance, and encouraging confederate-guide interaction. Our work furthers understanding of guides as a versatile method for VR accessibility and presents design recommendations for future guides.
\end{abstract}

%% file: 2Introduction.tex
\section{Introduction}
Social virtual reality (VR) applications, which allow multiple people to interact in virtual spaces, are among the most popular VR experiences. For example, VRChat regularly hosts 40,000 players per month \cite{vrchatmetrics}. However, the complex nature of social VR, where key visual information changes frequently, makes it particularly challenging for blind and low vision (BLV) people. Current social VR platforms do not have accessibility features to address these challenges.

Previous research on VR accessibility for BLV users has largely focused on adding spatial audio or haptic feedback to objects and actions, such as footstep sounds signifying avatar movement or haptic vibrations signifying collisions \cite{andrade2018echo, zhao2018cane, siu2020cane, zhang2020cane, dang2023opportunities}. For example, Dang et al. proposed a dual-track audio system that allows BLV users to hear spatial audio alongside concurrent audio descriptions of important actions during VR concerts \cite{dang2023opportunities}. However, this research exclusively targets single-user VR experiences. These solutions would likely be sensorially overwhelming and disruptive in the dynamic environments of social VR, requiring a fundamentally different, high-level approach. This highlights the need for a solution that can provide real-time, contextual information about scenes, rather than relying solely on low-level sensory feedback.

Recently, some researchers have addressed social VR accessibility using guiding interactions as a template \cite{collins2023gaze, wieland2022nvc, ji2022vrbubble, collins2023guide}. Previously, we proposed a sighted guide framework for BLV users in social VR, inspired by physical-world sighted guides, where a BLV person holds a sighted person's elbow for navigation support while the sighted person walks slightly ahead and to the side of them \cite{collins2023guide}. In this framework, a human sighted guide joins a BLV user to assist them in exploring virtual environments. We showed that virtual guides were a promising, easy-to-learn accessibility approach, allowing BLV users to successfully navigate virtual environments with a technique already familiar to them from the physical world. While highly effective, human guides suffer from major limitations: they are not always available, and BLV users may prefer using automated tools to maintain their independence \cite{collins2023guide, gonzalez2024scenedesc}. To overcome these challenges, the field requires an effective, on-demand guidance system that preserves a user’s autonomy, such as AI-driven tools.

To address challenges with human guides, we presented an AI guide for social VR in follow-up work \cite{collins2024ai}. We designed a guide that used a large language model (LLM) to answer users’ visual questions and provide navigation support inside VR. The guide also offered six “personas,” or combinations of appearances and mannerisms, to allow for a customizable guidance experience. Crucially, we have not yet evaluated this system's efficacy via a user study. This lack of empirical data leaves critical questions about BLV people's experience with an AI guide unanswered: Would BLV users be able to effectively utilize it? What behaviors would they exhibit toward the AI guide? Are AI-generated descriptions of social VR useful to BLV people in practice? Understanding the answers to these questions is necessary for designing an effective AI guide.

To address this need, we studied our previously developed AI guide’s \cite{collins2024ai} use in social VR environments with BLV users. In this work, our research questions are broad:  

\begin{itemize}
    \item \textit{RQ1: How is an AI guide utilized by BLV people in social virtual reality?}
    \item \textit{RQ2: How do BLV people behave towards an AI guide in social virtual reality?}
\end{itemize}

To answer these questions, we conducted a study with 16 BLV participants who utilized the AI guide in social VR environments modeled after parks. We selected two of the guide's personas—a dog and a robot—for them to experience. Participants completed two tasks in VR: 1) using the guide to become familiar with the layout of each park, and 2) leading park tours for confederates who joined the VR scene. This second task required participants to balance interaction with the environment, other users, and the AI guide. After all tasks were completed, we interviewed participants about their experience.

We found that participants were able to use the guide to explore the parks and share accurate information about their features with confederates. However, participants’ interactions with the guide varied significantly based on context. For example, participants treated the guide as a tool to address their visual needs when using it alone, but began role-playing with it around confederates. Many participants also seemed less satisfied with the guide’s descriptions when with confederates, rephrasing its descriptions with their own imagined details, such as invented backstories for avatars in the parks. We also observed that some participants treated the guide as a source of general knowledge, such as asking it for the correct terms of park features. Finally, we discuss how the observed use of the AI guide differs from use of human guides and identify opportunities for studying the use of AI guides over time.

We implemented the first AI guide based on feedback from and tested with BLV users. Through this, we contribute the first findings on an AI guide’s effectiveness with assisting BLV people in social VR and how they react to its assistance. This work offers a critical foundation for designing future AI-powered VR accessibility tools that are genuinely useful and empowering.

%% file: 3RelatedWork.tex
\section{Related Work}

Below, we review work on enhancing the accessibility of social VR environments and supporting BLV individuals with guides and visual interpretation in the physical world.

\subsection{Enhancing the Accessibility of Social VR Environments}

Social virtual environments are typically complex, requiring users to navigate surroundings, interpret avatar movements, and discern nonverbal cues. Researchers have explored many approaches to making these environments accessible, often by adding sonification or haptic effects to objects and users’ actions \cite{ji2022vrbubble, balasubramanian2023sceneweaver, jain2015hmd, goncalves2023navaware, killough2025vrsight, killough2025vrsightdemo}. For instance, systems like Ji et al.’s VRBubble \cite{ji2022vrbubble} and those by Killough et al. \cite{killough2025vrsight, killough2025vrsightdemo} use spatial audio feedback to convey the number, proximity, and location of avatars and objects in VR scenes. Others have focused on translating nonverbal cues, like eye contact and head nodding, into audio signals and haptic vibrations \cite{wieland2022nvc, wieland2023gaze, segal2024socialcues, dang2023music, collins2023gaze, jung2024accessible}. While crucial for low-level perception, these discrete sensory solutions often become sensorially overwhelming in the dense landscapes of social VR, and they fail to provide the high-level information necessary for task completion and social fluency. Our work seeks to address these issues by providing a guide tailored to provide such high-level output for complex social environments.

Enhancing the accessibility of others’ location and behavior addresses some aspects of VR experiences, but it is equally important to consider how disabled users wish to represent themselves during these experiences. To this end, researchers have explored how disabled people wish to present in social virtual worlds \cite{gualano2023invisible, gualano2024looking, zhang2023signifier, zhang2022selfpres, mack2023avatarrep}. Zhang et al. highlighted the lack of disability representation in VR avatar customization, with some disabilities being excluded from menus and others given only one option, such as a hearing aid for d/Deaf people \cite{zhang2022selfpres}. Gualano et al. explored how people with invisible disabilities like neurodivergence or chronic pain wish to represent their disabilities, finding that symbolic representations like social batteries were desired \cite{gualano2023invisible}. Our work complements this stream of research by exploring how an AI guide can provide assistance while offering new forms of disability representation through its own embodiments.

Among efforts to enhance social VR accessibility, our research focuses on exploring virtual guides to support BLV users. In our prior work by Collins and Jung et al., we introduced a framework featuring a remote human assistant who served as both a conversational partner and a tool for navigation and visual interpretation \cite{collins2023guide}. In this study, BLV users explored virtual parks with the assistance of this human guide. This human guide framework demonstrated high effectiveness in helping BLV users explore virtual parks. However, this approach is fundamentally limited by the availability of human assistants, users’ wish to be independent, and difficulty in matching guiding styles to individual preferences.

To address the limitations of human guides, our subsequent work by Collins et al. described the design of an AI guide with six combinations of speech characteristics and appearances called “personas.” \cite{collins2024ai} These personas offer varying levels of social interaction and fantastical embodiment, including a friendly guide dog and blunt white cane, catering to diverse user needs and providing new avenues for disability representation in social VR. However, this work also has several limitations. First, this guide was not tested with BLV users in social VR, limiting our understanding of how BLV people utilize it. Second, while the guide was co-designed with BLV members of our research team, it lacked broader design input from a general BLV audience. Finally, we did not provide explicit details about the LLM prompt used to enable our guide, which is crucial to others’ ability to create similar AI-powered assistants for BLV users.

Thus, we improved the initial design of our AI guide prototype, focusing on three key personas. We incorporated key modifications based on new pilot studies and an updated design process, and shared our full AI prompt as supplemental material. Our current work moves beyond our prior design focus by providing the first empirical evidence of how BLV individuals utilize and interact with an AI guide in social VR. By incorporating their feedback, we identified opportunities to build reliable, user-centered guides informed by the unique ways BLV individuals interact with AI assistance in VR.

\subsection{Supporting BLV People with Guides in the Physical World}

Guides are a widespread tool used to support BLV people in the physical world. Researchers have examined the use of guides by BLV people for decades \cite{kayukawa2020blindpilot, garaj2003pedestrian, guerreiro2019cabot, lannan2019campus, jagt2023familiar}. A “guide” can be any assistant providing physical support to navigate from one area to another. This includes dog guides (commonly referred to as “guide dogs”), robots, and sighted human guides (assistants offering physical navigation assistance). 

Sighted human guides and guide dogs have been widely studied for their dual roles as companions and navigators \cite{ball2022runners, miner2001dogguide, vincenzi2021interdependence, small2015interconnecting, macpherson2016guiding}. For instance, Small found via autoethnographic entries and questionnaires that BLV tourists fostered companionship with sighted tourists who guided them, allowing them to enjoy visual aspects of tourism, like artwork, through informal socializing \cite{small2015interconnecting}. Miner et al. conducted open-ended interviews with eight BLV people about working with guide dogs, finding that guide dogs increased their handlers’ social confidence and acted as icebreakers in social spaces as well as met navigational needs \cite{miner2001dogguide}.

Although the companionship of sighted guides and guide dogs is valued, less-social guidance robots have also emerged as a topic of interest among HCI researchers \cite{kayukawa2020blindpilot, guerreiro2019cabot, azenkot2016servicerobots, kayukawa2023museum, bonani2018robots, hwang2024lessons, hwang2024towards}. For example, Hwang et al. found that with the increased availability of robots, robot guides could serve as less expensive alternatives to guide dogs if they borrow best practices from guide dogs \cite{hwang2024lessons}. In follow-up work, Hwang et al. conducted interviews and observed guidance sessions with five guide dog trainers and 23 guide dog handlers to identify these practices to inform future robot guides \cite{hwang2024towards}.

Given the proven utility of guides in complex situations, we aimed to develop virtual guides for VR by incorporating established best practices from their physical-world counterparts.

\subsection{Supporting BLV People with Visual Interpretation in the Physical World}

Another key tool to support BLV people in the physical world are the well-studied visual interpretation services \cite{gonzalez2024scenedesc, gonzalez2025towards, garaj2003pedestrian, lannan2019campus, bigham2010vizwiz}. These services allow BLV people to capture images or video of their surroundings, often via camera-based phone applications, and receive verbal descriptions of the captured media from human or AI interpreters.

One well-known visual interpretation application is VizWiz, an application that connects BLV users to remote sighted workers who answer queries about provided images \cite{bigham2010vizwiz}. Similar human-powered visual interpretation services that have been studied include Aira and BeMyEyes \cite{garaj2003pedestrian, lannan2019campus}. For instance, Lannan interviewed nine BLV college students about how they used Aira, finding that the application helped them engage in group activities as well as navigate their campuses \cite{lannan2019campus}.

In recent years, some researchers have begun focusing on the use of AI-powered visual interpretation, driven by the increased availability and quality of descriptions generated by LLMs \cite{gonzalez2024scenedesc, gonzalez2025towards, lin2025viptour}. For example, Gonzalez and Collins et al. conducted a two-week diary study of an LLM-powered interpretation application, followed by interviews with 16 BLV people. They found BLV people preferred AI in certain use cases over human interpretation. These included reading financial documents and taking pictures in private areas like bathrooms since AI posed fewer privacy concerns \cite{gonzalez2024scenedesc}. Lin et al. explored using an AI-driven system to help 30 BLV participants navigate unfamiliar environments, finding that BLV people prioritized obstacle avoidance, wayfinding, and understanding environmental aesthetics distilled in a concise, personalized manner \cite{lin2025viptour}.

Given the many uses of visual interpretation tools, we aimed to create an AI guide that is able to respond to all types of queries, and can provide reliable enough information that users feel comfortable utilizing it even for private scenarios.

%% file: 4Methods.tex
\section{Methods}

\input{virtual_environments.tex}

We refined the AI guide prototype from our previous design \cite{collins2024ai}, implementing specific changes (such as specific personas and re-engineered prompts) to study its use. We recruited BLV participants and observed them performing individual and social tasks in virtual environments using the guide. Finally, we conducted interviews about participants’ VR experience.

\subsection{Participants}

We recruited 16 participants (see Table \ref{tab: Participants}) through a partner organization, the LightHouse for the Blind and Visually Impaired in San Francisco, which provides services to BLV individuals. Participants were eligible for our study if they identified as blind or low vision, were at least 18 years old, could travel for an in-person study, and met the Meta Quest Health and Safety Guidelines \cite{guidelines}. Each participant was compensated \$100. Participants’ ages ranged from 22 to 75 (SD = 14.23, 7 women, 9 men). Eight participants identified as low vision and eight as blind. All procedures were approved by our university’s Institutional Review Board.

\input{participants.tex}

\subsection{Procedure} \label{sec: Procedure}

We conducted a single, 90-minute study session with each participant at a conference room provided by LightHouse. Each session comprised three parts: a VR tutorial, two rounds of VR tasks, and a semi-structured interview. During the VR tutorial, participants explored the AI guide’s features in its human persona. Participants then completed the VR tasks using the dog and the robot personas (see section \ref{sec: Prototype Design and Implementation} for explanation of personas). These tasks were counterbalanced so that each participant experienced both personas as well as different environments (see Table \ref{tab: Counterbalancing Procedures} for detailed counterbalancing and study procedures).

\input{procedure.tex}

The tutorial began in a simple square room filled with primitive 3D objects. This introduction allowed participants to become familiar with the VR platform and the AI guide controls. Participants learned to move around the room, pick up objects, and call their guide. The tutorial concluded once participants felt comfortable with these basic interactions.

Participants then proceeded to the first of two VR tasks: exploring a virtual park with the AI guide. The goal of this task was to observe use of a guide in \textit{individual} settings. Participants were placed in one of two virtual environments modeled after parks from our study on human guides (see section \ref{sec: Prototype Design and Implementation}) \cite{collins2023guide}. Participants were asked to familiarize themselves with key landmarks (e.g., gazebos, a river). This task ended after ten minutes or once the participant indicated they were comfortable with the park contents.

The second task required participants to give a tour of the same park to two joining researchers acting as confederates. This task was designed to observe guide use in \textit{social} settings, where participants had to balance their use of the guide with social interaction. Participants were instructed to answer any questions from the confederates and lead them to at least two park landmarks. This task concluded after fifteen minutes or once they had completed the tour. Participants then repeated both tasks in a different park with a new guide persona.

Throughout the entire study, confederates used human avatars matching their preferred gender. To standardize confederate behavior, senior researchers on the team developed behavioral guidelines (included in supplementary materials). For example, the guidelines instructed confederates to pretend to be sighted participants of a different study who were unfamiliar with the guide. We implemented this to avoid influencing participants’ guide usage if they believed the confederates were experienced with it. In addition, the guidelines provided examples of questions and topics to discuss (e.g., the guide, the parks, hobbies) to ensure natural interaction. They also included instructions for movement throughout the study, such as pre-determined rendezvous points with participants, best practices for following participants, and examples of natural body language or gestures to use.

We concluded our study with a 30-minute semi-structured interview. Participants responded to Likert-scale statements (rated 1 to 5; see Figure \ref{fig: Likert Guide Ratings}) and answered open-ended questions across nine categories: usability, usefulness, joy of use, social comfort, scene understanding, object perception, navigation, appearance, and areas for improvement. Example questions included:

\begin{itemize}
	\item \textbf{“Which of the guide personas did you prefer using? Why?”}
	\item \textbf{“Were there any tasks you felt the guide struggled to assist you with?”}
	\item \textbf{“How effective were the guide’s descriptions of the scenes you were in?”}
\item \textbf{“How do you believe the guide impacted others’ perceptions of you?”}
\end{itemize}

\subsection{Prototype Design and Implementation} \label{sec: Prototype Design and Implementation}

We refined our original design of an AI guide \cite{collins2024ai}, incorporating specific changes informed by pilot studies in our current work.

Our refined AI guide offered three primary capabilities of the original guide: (1) providing \textbf{visual descriptions} of the environment, (2) assisting users in \textbf{moving their avatars} to specified locations or objects upon request, and (3) placing spatialized \textbf{audio beacons} on objects to aid in orientation. Users could activate or mute the guide using the secondary buttons on their VR controllers. They interacted with the guide via specific (e.g. “Take me to the Northern Fountain”) or vague (e.g., “What is this bluish-gray blob in front of me?”) voice commands.

\input{personas.tex}

In our original work, we also proposed six guide “personas,” or combinations of different appearances and personalities, to provide varied guidance experiences. While a comprehensive evaluation of all personas is an area for future research, our objective was to evaluate the AI guide's core functionality, so we implemented only three within our refined guide: the dog, the robot, and the human (see Table \ref{tab: Guide Personas}). We selected the dog because it is a familiar physical-world guide form that also serves as a disability signifier. We chose the robot to be a fantastical form representative of the unique experiences possible in VR. Finally, we used the human for the tutorial, as its similarity to a human sighted guide facilitated an easier transition to the VR framework. We named all personas "Giddy," unlike the original AI guide which had no name. This decision—intended to give the guide a simple, short, and androgynous name—was based on suggestions from BLV participants in pilot studies informing our updated AI guide design.

To replicate the initial guidance experience from our human guide study, we re-utilized two virtual park environments from that study \cite{collins2023guide}. The first park had a calm environment with a square pathway and a river, while the second was more lively, featuring a complex pathway and a group of dancers. We also included a tutorial environment with simplistic objects for training purposes (see Figure \ref{fig: Virtual Environments}).

We implemented our VR prototype in Unity, following our original specifications from Collins et al. \cite{collins2024ai}. Navigation relied on Unity's NavMesh and NavAgents pathfinding components, and user queries were processed via OpenAI APIs. We used OpenAI's Speech-to-Text model, Whisper, to convert verbal queries to text, which was then sent to GPT-4 along with environment screenshots captured at the time of the query and hosted in a temporary image server. The guide then sorted these queries into five categories with different output types and used descriptors to embody the intended persona. To assess system responsiveness, we defined and recorded the guide's response time as the duration from the end of the user's speech input to the start of the guide's audio output. Finally, we built our prototype on a Meta Quest 2. Utilizing a visually-dominant VR headset presents a potential tradeoff of increasing cognitive load or physical discomfort for BLV users experiencing it as an audio-centric system. However, we chose this hardware to ensure the prototype's compatibility with standard, widely-available commercial VR ecosystems and consumer VR experiences.

Prior to deployment, we investigated several Speech-to-Text models for their robustness, including Google's Speech-to-Text API and Microsoft’s Azure AI Speech. We selected OpenAI’s Whisper due to its integration within the same API services as GPT-4 (also from OpenAI). This unified architecture allowed for more seamless data transfer between our STT and LLM endpoints, resulting in lower operational costs and lower end-to-end latency during pilot testing. Ultimately, Whisper performed adequately for our general population.

In addition to the basic prompt structure from our prior work, we developed a new prompt section based on pilot study feedback. This section instructed GPT-4 to write responses specifically for a blind user, incorporating best practices identified by our BLV pilot participants. These practices included using distances, cardinal directions, and references to the user's position relative to known objects. We also limited responses to 200 words or less to prevent overwhelming the user with extraneous detail, a limitation not imposed in our prior work. Finally, the guide was instructed to structure its responses directly to the user, as sometimes, it would reply as if to an unknown third party. The full prompt is available in our supplemental materials.

\subsection{Data and Analysis}
We collected video recordings of the VR tasks and audio recordings of the semi-structured interviews. Four researchers then transcribed and coded these recordings using a mixed-methods approach.

First, three researchers coded a transcript with closed codes to: (1) tally the number of participant-guide interactions and the guide’s correct responses, and (2) note types of interactions between the participant and guide, following interaction categorizations from our human guide study \cite{collins2023guide}. Discrepancies were discussed and a consensus reached before dividing remaining transcripts for coding. The closed codes were analyzed separately to complement the qualitative findings.

For the qualitative data, two researchers coded a transcript with open codes to create an initial codebook, which was then applied to the remaining transcripts. This was followed by an inductive thematic analysis \cite{clarke2017thematic, braun2019reflecting} using affinity diagramming to identify emerging themes.

As part of coding interactions, researchers needed to classify the tones in which participants addressed the guide. We initially used tone categories from our previous work with the human guide (utilitarian, friendly, respectful, apologetic, and uncertain) \cite{collins2023guide}. However, after finding no instances of behaviors like apologizing or deferring expertise, we reduced the categories to utilitarian, friendly, and polite. We reclassified "respectful" to "polite" based on coder consensus that participants were not offering consistent respect, but using occasional polite markers like "please" and "thank you." We classified any queries where participants used such phrases as “polite.”

To classify friendly and utilitarian speech, we deferred to external sources on social dynamics and linguistics. We defined “friendly” queries based on behaviors that foster social connection \cite{angelo2015, goodwin1990}. For example, studies on social dynamics note that using a person's name or offering greetings demonstrates friendliness or connection \cite{angelo2015, goodwin1990}. Thus, we classified queries where participants used the guide’s name, a nickname, or greeted the guide before asking questions as “friendly” interactions. Conversely, a “utilitarian” tone was defined by language that ordered the guide to perform actions without social markers. Following linguistic classifications of imperative language \cite{kaufmann2011}, we identified utilitarian queries by their sentence structures, which manifested as orders, commands, or requests with no polite or friendly alterations.

%% file: virtual_environments.tex
\begin{figure*}[t]
\centering
\includegraphics[width=350pt]{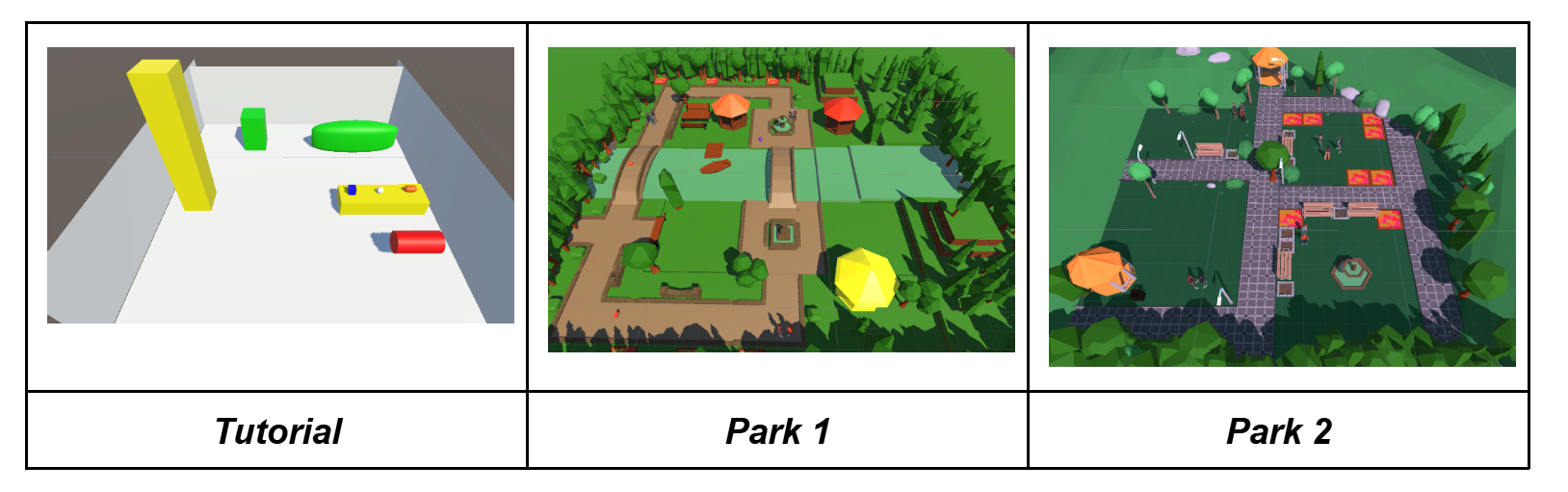}
\caption{\label{fig: Virtual Environments} The three virtual environments used in the study. From left to right, the Tutorial contains various simplistic 3D objects such as yellow and green cubes, Park 1 contains a large river and several colorful gazebos, and Park 2 contains colorful flower beds and a series of benches for resting throughout the park.}
    \Description{A table with three images of the virtual environments used in the study. On the far left is an image of the tutorial room, a simple white room with various geometrical shapes, including a yellow, horizontal cube acting as a table with three interactable shapes resting on it. In the middle is a screenshot of the first park with trees scattered about and a river cutting through, populated by stone walkways and bridges, two water fountains, three gazebos, and three groups of avatars. On the right is a screenshot of the second park, a brightly-lit park with stone tile walkways around a few square patches of grass, two gazebos, and a group of avatars gathered for a dance party.
}
\end{figure*}

%% file: participants.tex
\begin{table*}[h!]
    \footnotesize 
    \centering
    \caption{\label{tab: Participants} Participant demographics. F = female, M = male. Participant information is self-reported. NR indicates that participants chose not to report that information.} 
    \begin{tabular}{p{1.4cm}p{0.3cm}p{1.0cm}p{2.0cm}p{1.5cm}p{2.0cm}p{2.0cm}p{4.0cm}}
    \toprule
      \textbf{Pseudonym}&  \textbf{Age}&  \textbf{Gender}&  \textbf{Visual Condition}&  \textbf{Etiology}&  \textbf{Visual Acuity}&   \textbf{Visual Field}&  \textbf{Navigation Tools Used} \\
     \midrule 
    
   \textbf{Niya}&   45&   F&  Blind; No Light Perception&  NR&    NR&     NR&  White Cane; Human Guide; Navigation Apps; Tactile Maps\\ 
    \midrule
    
    \textbf{Anika}&   46& F&  Low Vision&   Optic Atrophy&    R: 20/400, L: 20/200&     Vision in upper field of left eye, right eye blurry&  White Cane; Navigation Apps\\ 
    \midrule

     \textbf{Teo} & 51 & M&  Low Vision&   Cortical Impairment&    R: 20/400, L: 20/400&     NR&  White Cane; Navigation Apps; GPS Devices\\ 
    \midrule
    
      \textbf{Nikolai}&    34&    M& Partially Sighted; Legally Blind; Night Blindness; Central Vision Loss&   NR&    NR&     NR&  Navigation Apps\\ 
    \midrule
    
      \textbf{Sara}&    22& F&    Legally Blind&     Dual Optic Coloboma&    NR&     NR&  White Cane; Navigation Apps\\ 
    \midrule

      \textbf{Xiran}&    38& F&    Legally Blind; Night Blindness; Peripheral Vision Loss &    NR&    NR&     NR&  White Cane; Human Guide; Navigation Apps\\ 
    \midrule
    
      \textbf{Yuki}&    37& M&    Blind; No Light Perception&     NR&    NR&     NR&  White Cane; Navigation Apps\\ 
    \midrule
    
      \textbf{Maritza}&    75& F&   Blind; No Light Perception&     NR&    NR&     NR&  White Cane; Guide Dog; Human Guide; Navigation Apps; GPS Devices\\ 
    \midrule
    
      \textbf{Aiden}&    45& M&   Blind; Light Perception Only; Legally Blind&     NR&    NR&     No vision in left eye, light detection in right eye&  White Cane; Human Guide; Navigation Apps; GPS Devices\\ 
    \midrule
    
      \textbf{Len}&    29& M&  Low Vision; Legally Blind; Night Blindness; Peripheral Vision Loss&    Retinitis Pigmentosa&    NR&     NR&  White Cane; Human Guide; Navigation Apps\\ 
    \midrule
    
      \textbf{Siobhan}&    55& F&  Blind; No Light Perception&   Retinopathy of Prematurity&    NR&     NR&  White Cane; Guide Dog; Navigation Apps; Tactile Maps; GPS Devices; AR Glasses\\ 
    \midrule
    
      \textbf{Isaiah}&    54& M&  Partially Sighted&    NR&    R: 20/200, L: 20/150&     NR&  White Cane; Human Guide; Navigation Apps\\ 
    \midrule
    
      \textbf{Li}&    55& M&  Low Vision; Legally Blind; Peripheral Vision Loss&    Diabetic Retinopathy&    NR&     NR&  White Cane; Human Guide; Navigation Apps\\ 
    \midrule

      \textbf{Jamal}&    67& M&  Low Vision&    NR&    R: None, L: 20/200&     No vision in right eye, minimal field in left eye&  Navigation Apps\\ 
    \midrule

      \textbf{Yadriel}&    27& M&  Low Vision; Color Blindness&    Autosomal Optic Atrophy&    R: 20/300, L: 20/400&     Normal visual field&  Do not use any\\ 
    \midrule

      \textbf{Jiwoo}&    28& M&  Low Vision; Partially Sighted; Legally Blind; Color Blindness&    NR&    NR&     NR&  Navigation Apps\\ 
    
     \bottomrule
    \end{tabular}
    \Description{Information about our participant demographics. Contains all textual content.}
\end{table*}

%% file: procedure.tex
\begin{table*}[htbp]
\centering
\caption{\label{tab: Counterbalancing Procedures}The counterbalanced study conditions (e.g., duration, park, persona) across all groups of participants. Group 1 = Niya, Anika, Teo, Nikolai; Group 2 = Sara, Xiran, Yuki, Maritza; Group 3 = Aiden, Len, Siobhan, Isaiah; Group 4 = Li, Jamal, Yadriel, Jiwoo. Participants completed two tasks per study phase (i.e., one individual and one social task).}
\label{tab:study_conditions}
\setlength{\tabcolsep}{5pt} % Adjust column separation for better fit
\begin{tabular}{cccccc}
\toprule
\textbf{Group} & \textbf{Phase} & \textbf{Task Type} & \textbf{Duration} & \textbf{Park} & \textbf{Persona} \\
\textbf{(N = 4 participants per group)} & & & & \textbf{Assignment} & \textbf{Assignment} \\
\midrule
\multirow{4}{*}{\textbf{1}} & \multirow{2}{*}{Round 1} & Individual Task (Exploration) & 10 minutes & \multirow{2}{*}{Park 1} & \multirow{2}{*}{Dog} \\
 & & Social Task (Tour) & 15 minutes & & \\
\cmidrule{2-6}
 & \multirow{2}{*}{Round 2} & Individual Task (Exploration) & 10 minutes & \multirow{2}{*}{Park 2} & \multirow{2}{*}{Robot} \\
 & & Social Task (Tour) & 15 minutes & & \\
\midrule
\multirow{4}{*}{\textbf{2}} & \multirow{2}{*}{Round 1} & Individual Task (Exploration) & 10 minutes & \multirow{2}{*}{Park 1} & \multirow{2}{*}{Dog} \\
 & & Social Task (Tour) & 15 minutes & & \\
\cmidrule{2-6}
 & \multirow{2}{*}{Round 2} & Individual Task (Exploration) & 10 minutes & \multirow{2}{*}{Park 2} & \multirow{2}{*}{Robot} \\
 & & Social Task (Tour) & 15 minutes & & \\
\midrule
\multirow{4}{*}{\textbf{3}} & \multirow{2}{*}{Round 1} & Individual Task (Exploration) & 10 minutes & \multirow{2}{*}{Park 1} & \multirow{2}{*}{Robot} \\
 & & Social Task (Tour) & 15 minutes & & \\
\cmidrule{2-6}
 & \multirow{2}{*}{Round 2} & Individual Task (Exploration) & 10 minutes & \multirow{2}{*}{Park 2} & \multirow{2}{*}{Dog} \\
 & & Social Task (Tour) & 15 minutes & & \\
\midrule
\multirow{4}{*}{\textbf{4}} & \multirow{2}{*}{Round 1} & Individual Task (Exploration) & 10 minutes & \multirow{2}{*}{Park 1} & \multirow{2}{*}{Robot} \\
 & & Social Task (Tour) & 15 minutes & & \\
\cmidrule{2-6}
 & \multirow{2}{*}{Round 2} & Individual Task (Exploration) & 10 minutes & \multirow{2}{*}{Park 2} & \multirow{2}{*}{Dog} \\
 & & Social Task (Tour) & 15 minutes & & \\
\bottomrule
\end{tabular}
\end{table*}

%% file: personas.tex
\begin{table*}[t]
            \small
            \renewcommand{\arraystretch}{1.7}
            \begin{center}
            \caption{\label{tab: Guide Personas} Our guide’s three personas: the human, the guide dog, and the robot. The voices mentioned for each persona (Alloy, Nova, and Onyx) are OpenAI’s default TTS voices \cite{openai-tts}. Audio samples of each voice can be found at the cited link. The personality prompts were sent to GPT-4 to shape the guide's responses (see supplemental materials). All columns are based on our previous designs from Collins et al. \cite{collins2024ai}, except for Speech Sample, which was produced by our current implementation of the AI guide.}
                \begin{tabular}{|P{1cm} | M{2cm} | P{2cm} | P{2cm} | P{2cm} | P{2cm} |P{3.2cm} |}
                    \hline
                    \textbf{Persona} & \textbf{Guide Avatar} & \textbf{Guide Description} &
                    \textbf{Guide Movement} &
                    \textbf{Voice + Speech Characteristics} &
                    \textbf{Personality Prompt} & \textbf{Speech Sample} \\ \hline
                    \textbf{Human} & \raisebox{-\totalheight}{\includegraphics[width=1.8cm, height=2.3cm]{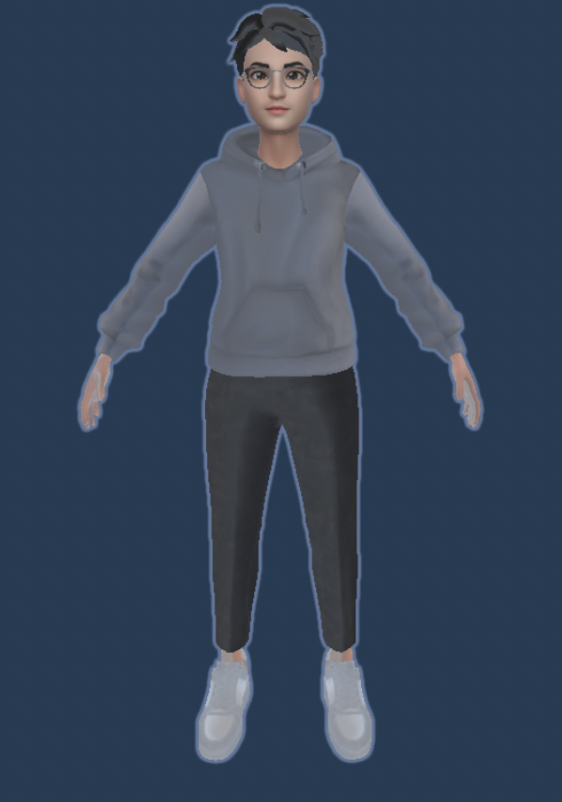}} & An androgynous human with short black hair and glasses, wearing a gray hoodie and black pants. & Follows behind the user from an arm’s length away. Footstep sound effects indicate movement. & \textit{Alloy,} androgynous voice, medium formality (some jokes, casual speech), averages 101.2 words per response & Warm, friendly, but still professional sighted guide & \textit{The scene around you is a simple, geometric landscape composed of distinctive objects. To your right stands the Tall Building, a towering yellow cube that rises high above the rest…} \\
                    \hline
                    \textbf{Guide Dog} & \raisebox{-\totalheight}{\includegraphics[width=2cm, height=2cm]{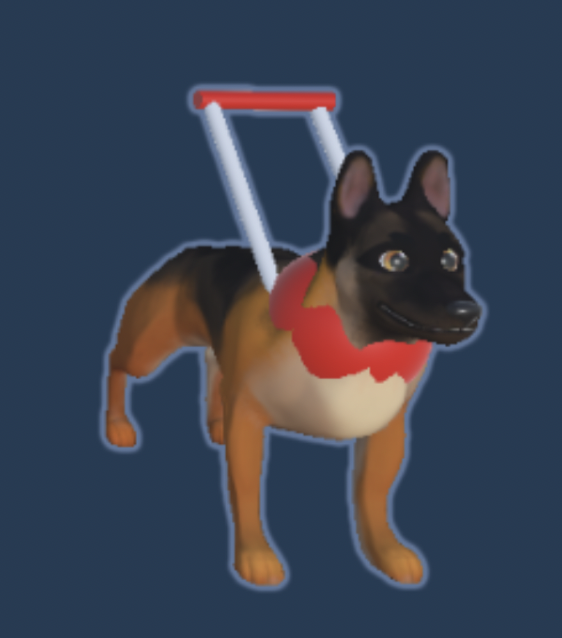}} & German-shepherd with black and brown fur, wearing a bright-red guide dog harness. & Walks beside the user on their left side. The sounds of panting and paws scratching the floor indicate movement. & \textit{Nova,} feminine voice, low formality (uses slang, jokes, and casual speech), averages 185.4 words per response & Very friendly, excited companion who is eager to please who you’re talking to & \textit{What an exciting world we're in! To your right stands the impressive Tall Building, reaching up high with its yellow facade…} \\
                    \hline
                    \textbf{Robot} & \raisebox{-\totalheight}{\includegraphics[width=1.8cm, height=2.3cm]{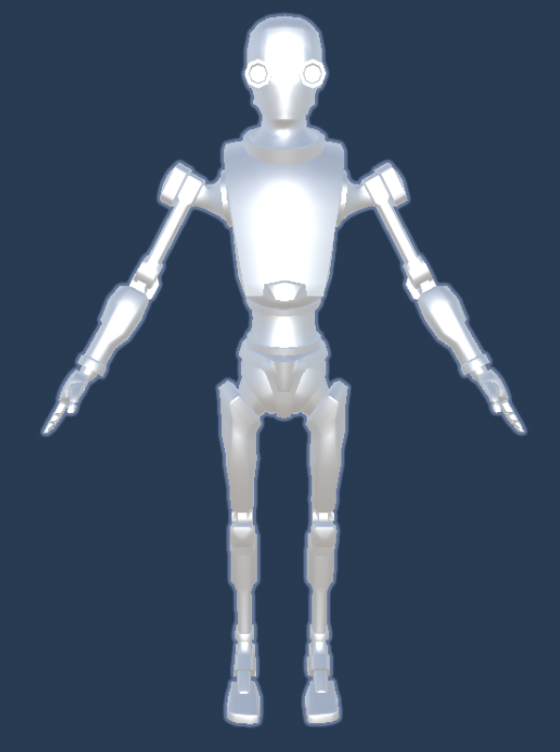}} & A shiny, silver robot that is humanoid in stature, with a masculine frame. & Follows behind the user from an arm’s length away. Creaking machinery sound effects indicate movement. & \textit{Onyx,} masculine voice, high formality (speaks entirely in proper sentences, no casual speech), averages 91.2 words per response & Formal and assertive assistant who talks like a robot & \textit{Affirmative. The area around you is composed of geometric shapes and structures. Directly to your right lies the Tall Building…} \\
                    \hline
                \end{tabular}
                \newline\newline
                \Description{Information about three different guide personas. Contains all textual content, aside from the three images of the persona's avatars in the column "Guide Avatar." Each avatar is described in the following column "Guide Description."}
            \end{center}
        \end{table*}

%% file: 5Findings.tex
\section{Findings}

We report participants’ perceptions of the AI guide, the guide’s accuracy across tasks, and how participants behaved towards the guide in solo and social settings.

\subsection{Overview}
All participants explored both parks and attempted to provide tours. While describing the parks to the confederates, all participants provided accurate information about the parks, demonstrating that they had learned the park contents. However, providing a tour proved more challenging: 10 out of 16 participants met the requirement of guiding their group to two landmarks, three guided them to more than two, and six made it to one. Participants who struggled with this task often forgot the names of landmarks and that they could give the guide descriptions of desired landmarks instead of names to trigger guidance.

Other participants struggled due to inaccuracies made by the guide. While the guide responded correctly to a total of 301 out of 476 queries (63.2\%), factors such as misinterpreting participant accents or receiving incomplete queries from participants reduced guide accuracy. When responding to requests, the guide’s average response time was 6.1 seconds for guidance requests, 11.2 seconds for visual questions, and 6.8 seconds for interpersonal queries.  We categorized queries into seven request types: navigation (e.g., “Take me to the fountain”), visual description (e.g., “Describe the structure in front of me”), clarification (e.g., “Could you repeat what you said? I didn’t understand.”), audio beacon (e.g., “Put an audio beacon on the northern gazebo”), social interaction (e.g., “Say ‘hi’ to my friends here, Giddy”), confirmation (e.g., “I’m looking at a patch of flowers, right?”), and auditory description (“What’s making that watery sound?”). Of these types, navigation and visual interpretation were most common, making up 60.4\% and 31.4\% of requests, respectively.

When conversing with the guide, participants typically spoke with a utilitarian tone (see Figure \ref{fig: request_tone}), with 74.6\% of interactions being straightforward and commanding (e.g., “Take me to the fountain”). 15.8\% were polite (e.g., “Could you please take me to the fountain?”) and 9.6\% of remaining interactions were friendly (e.g., “Hey buddy, let’s go to the fountain”).

Though we did not explicitly compare participants’ experience with the two guide personas they used–the guide dog and robot–we did note differences in how they initially reacted to each. The dog persona was generally thought of as “cuter” (Anika, Niya) or more “empowering” (Aiden, Nikolai, Xiran), due to it being a symbol of blindness. Participants also demonstrated different behaviors towards the guide depending on its persona, which we detail in section \ref{sec: Behavior Towards the Guide}.

\input{request_type.tex}
\input{request_tone.tex}

\subsection{Participants’ Ratings of the Guide and their VR Experience}

We present participants’ assessments of the guide based on Likert scale scores reported in Figure \ref{fig: request_tone}. We do not include feedback from three participants (Li, Yadriel, Jiwoo) who did not use the guide. However, their experiences are described in section \ref{sec: Completing Tasks Without The Guide}.

\input{likert_responses.tex}

\textbf{Usability}. Four participants somewhat or strongly agreed that it was easy to work with the guide to complete tasks, noting that the controls were straightforward and the guide generally understood their queries (mean 3.2, SD = 1.1). For instance, Nikolai appreciated the ability to activate the guide without needing to know its exact location. Conversely, nine participants disagreed or felt neutral about the guide’s usability, reporting difficulties with controls.

Some participants (Niya, Maritza, Siobhan) found the guide's inconsistent response style challenging. Though inconsistency is a common quality of AI-generated content, where the AI changes how it replies so as to sound more natural, this quality was not helpful in a guidance context, especially for participants used to the consistent manner of communication typical of human guides. Siobhan remarked that she wished “the directions maybe, could be consistent each time, so that they are not quite as abstract, and they don't vary quite as significantly every single time.”

Other participants (Xiran, Aiden, Len, Jamal) expressed that the guide struggled to comprehend nuanced instructions. Xiran wanted the guide to bring her on a continuous tour around the park, moving from one landmark to another. However, when she attempted to prompt the guide for such navigation, it instead provided generic verbal directions like “walk straight” (Xiran). Similarly, Len tried to prompt the guide to go to a previously mentioned flower patch with slightly poetic language, saying: “Let’s go smell the flowers.” In response, the confused guide gave a description of the nearest flowers: “Smelling flowers sounds splendid! There are several vibrant flower patches around, each bursting with color and light…”

\textbf{Usefulness}. Despite usability challenges, nine participants found the guide useful, somewhat agreeing that it was able to address their commands and questions (mean 3.5, SD = 0.8). They commented that the guide felt “necessary” (Teo) to experience VR, and that it allowed them to get more information than “randomly exploring” (Anika) on their own would have.

\textbf{Joy of Use}. Nine participants enjoyed using the guide, somewhat or strongly agreeing that the guide was an enjoyable part of their experience (mean 4.1, SD = 1.0). Other participants were more neutral, feeling that it did not have to be enjoyable since it was an assistive tool (Jamal) or that enhancements were needed before they considered it enjoyable (Niya, Len, Siobhan). Len, for instance, expressed a desire for greater privacy when using the guide in social settings:

\begin{quote}
I would like the guide to just be something that only I can see and hear if I'm going to be using it in a social environment. I'd like to have [a] button…that mutes my chat with everyone else around me, so the only thing that's hearing me is my guide…For lack of a better term, [an] introvert guide setting where I could just be like, ‘Alright. This is \textit{my} guide for \textit{me} to see.’ …[that I could turn] on and off.
\end{quote}

\textbf{Social Comfort}. Nine participants reported feeling comfortable using the guide in front of other people (mean 3.7, SD = 1.3). They attributed their comfort to prior experiences using assistive devices. Aiden elaborated: 

\begin{quote}
I'm so used to… having a guide dog or a cane, and just explaining…to people ‘I'm blind…I've attended conferences with a guide dog before. So I think just having [the guide] is just going to be like, ‘Yeah, this is my assistant and he's helping me around.’ I think I'd be comfortable with that in pretty much any situation.
\end{quote}

However, four participants expressed discomfort, primarily due to the guide's visibility to others. Despite their familiarity with visible mobility aids in the physical world, they preferred to control how and when their disability was disclosed in VR. Len remarked:

\begin{quote}
I'm not going to want [others] to know that I'm blind or visually impaired right away. It kind of blows away the anonymity of the virtual world…and the way I want to present myself…I don't want to give away certain details about myself, [even] something minuscule like my visual impairment.
\end{quote}

Others (Niya, Sara) found the guide's lingering presence “awkward” (Niya) when not actively in use. Sara noted that this feeling was mitigated when using the dog guide, since the confederates seemed to like petting and interacting with it, but it still made her uncomfortable.

Finally, participants had mixed opinions on three key aspects of their ability to explore virtual scenes: scene understanding, object perception, and navigation.

\textbf{Scene Understanding}. Nine participants somewhat or strongly agreed that they had an accurate understanding of the scene (mean 3.6, SD = 1.3). Other participants struggled with this aspect, and some (Niya, Siobhan) wanted the guide to use cardinal directions more consistently, since switching between descriptions like “north and south…then bottom or left” (Niya) was confusing.

\textbf{Object Perception}. Nine participants somewhat or strongly agreed that it was easy for them to identify virtual objects (mean 3.6, SD = 1.1). Others found object perception more challenging, and some (Xiran, Len) commented that the guide did not improve their perception abilities. They explained that the guide’s descriptions of objects were “passable” (Xiran), but did not make them feel confident that they understood what they were looking at.

\textbf{Navigation}. Five participants somewhat or strongly agreed that it was easy to move around the parks (mean 3.1, SD = 1.3). Conversely, eight participants disagreed or felt neutral about their navigation abilities. Some participants (Aiden, Siobhan, Jamal) again referenced difficulties with the guide’s inconsistent descriptions, which did not help them build reliable “mental maps” (Siobhan, Jamal). Others (Niya, Anika, Teo, Maritza, Len, Jamal) experienced usability issues when attempting to navigate with the guide, such as forgetting the button to grab it.

\subsection{Completing Tasks with the Guide}
\label{Participant Strategies for Completing Tasks}

We report strategies for how participants approached the study tasks with the guide, in solo and social settings.

\subsubsection{Exploring Alone} \label{sec: Exploring Alone}

During solo park exploration, all participants used the guide to navigate and become familiar with park content, employing the guide in particular ways to achieve these goals. For navigation, all participants combined the guide’s abilities with their own, sometimes asking the guide to take them to different places, and other times walking or teleporting around the park themselves. To familiarize themselves with the park, nearly all participants (n = 10) posed brief questions to the guide when they needed information, such as asking for the names of objects as they encountered them. Several others (Anika, Teo, Nikolai, Xiran, Yuki, Maritza, Aiden, Len, Siobhan) used the guide to provide an overview of their environment, then chose individual details to explore further. For instance, after Yuki wandered off the path he had been following in one park, he called the guide to get his bearings again:

\begin{quote}
	\textbf{Yuki: Okay. And what's around me?}\\
Guide: You're surrounded by various things. The Red-Roofed Gazebo, the Pine Tree by the River, and more pine trees to your left. To your right is the Boat, and beyond that is the Dock–\\
Yuki: Oh, okay.\\
Guide: –To explore more of the scene, head onto the path beside the bench leading to Western Bridge. To your left, the Puffy Tree by the Bench provides some shade.\\
\textbf{Yuki: Take me to the bridge.}\\
Guide: Okay. Grab onto me, and I'll take you to Western Bridge.
\end{quote}

\subsubsection{Leading Others to Landmarks} \label{sec: Leading Others to Landmarks}

While exploring on their own, participants occasionally walked and teleported by themselves; however, during the group tours, they always used the guide to lead their group between landmarks. Participants used the guide to lead their groups in two ways. More commonly, they ordered the guide to take them to identified landmarks, and asked the confederates to follow them. Less commonly, they asked for an audio beacon or the guide misinterpreted their request and placed an audio beacon on a landmark. In this case, participants walked towards the sound of the beacon to lead their group, as Aiden did in the following exchange:

\begin{quote}
Aiden: Welcome to the park. I'm your tour guide for the day. So is there, is there anything you would like to see?\\
Confederate 1: I mean, if you have a favorite part of this park to take us to, then we can go there.\\
Aiden: Yeah, I kind of like the north platform there, where you can kind of see a lot of the different stuff around. So let's walk over there.\\
Confederate 1: Alright.\\
Confederate 2: That sounds cool.\\
\textbf{Aiden: Okay.} \textit{[To the guide]} \textbf{Walk me to the northern platform.}\\
Guide:  Very well. I will add an audio beacon to Northern Platform.\\
\textbf{Aiden:}  \textit{[To the confederates]} \textbf{Alright. So if you guys will follow me here, I'll walk up there.}
\end{quote}

\subsubsection{Filling Gaps in Knowledge} \label{sec: Filling Gaps in Knowledge}

Throughout the tours, participants used the guide to help them answer confederates’ questions. Typically, they first responded to questions based on what they had learned themselves during solo exploration. However, if participants needed additional information to answer fully, they called the guide. In the following conversation, Sara realized she had forgotten an important detail about the park as she was describing it:

\begin{quote}
\textbf{Sara:} Okay. So right now, we're on this dock. There's a boat in front of us. Nice little river running past the bridges. \textbf{There is one, two, so there's a, oh, boy! What is that called again? One second.} \textit{[To the guide]} \textbf{What is this structure?}\\
Guide: You are looking at the Orange-Roofed Gazebo.\\
\textbf{Sara: Oh, gazebo!} \textit{[To the confederates]} \textbf{I've never heard of that word before.} So there's an orange gazebo, a red gazebo, and a yellow gazebo.
\end{quote}

Maritza also used the guide to answer confederate questions, not to recall forgotten information, but to obtain details she had never learned:

\begin{quote}
Confederate 1: How big is this place?\\
\textbf{Maritza:} All right. Let's all go to the, oh. \textbf{We are [a] very short distance from everything.} \textit{[To the guide]} \textbf{How big is this park?}\\
Guide: How can I assist you in this park?\\
\textbf{Maritza: How big is this park?}\\
Guide: The park is moderately sized, featuring various points of interest, such as bridges spanning the river and multiple platforms. You'll also find benches, picnic tables, fountains, as well as designated areas with flowers and trees. There are groups of people congregating in different areas, adding life to the scene. What else would you like to know?
\end{quote}

However, many participants (Teo, Sara, Yuki, Len, Siobhan, Li, Jamal) were not satisfied with the guide’s answers and supplemented its responses. They summarized responses with their own interpretations or even added invented details. For example:

\begin{quote}
Yuki: Okay. Am I at the dance floor?\\
Guide: Hey there! I'm Giddy, your super guide for the rest of the park. How can I assist you on this adventure? If you find yourself toe tapping, it sounds like you're in need of dancing. You're currently surrounded by an active dance floor filled with the lively Dancing People. You might consider joining them! Nearby, there is also the serene calm of a gazebo–\\
\textit{[The participant cuts the guide off with the mute button and turns to the confederates.]}\\
\textbf{Yuki:} Thanks. Okay. \textbf{So where we are is at the dance floor. People come and will dance, or dance and maybe do aerobics like work out a little bit.}
\end{quote}

Here, Yuki added details such as that the park-goers do aerobics on the dance floor. Other participants added invented park names (“Kendrick Park”, Len) or made up stories about the relationships between avatars in the park (Sara).

\subsection{Completing Tasks Without The Guide} \label{sec: Completing Tasks Without The Guide}

Three participants (Li, Yadriel, Jiwoo) preferred to use their residual vision to complete tasks. Instead of using the guide, they leaned closer to objects, angled their heads to position objects better in their field of vision, or stuck to visually-distinct landmarks (e.g., following the gray stone path in the ground, moving from one patch of flowers to another). They answered confederates’ questions confidently and walked ahead of them as they led them around the park.

Although these participants did not need the guide in our study, they anticipated that the guide would be useful in visually complex scenarios. Li elaborated:

\begin{quote}
If the park was bigger…instead of [just] in a contained park, like on a city block, or something like that…I would want the guide…If it is a congested situation, it's always nice to have somebody on my left side that I can kind of follow and maneuver through [especially] if it's busy, lots of people quarreling.
\end{quote}

Yadriel also commented that larger environments like city layouts would encourage him to use the guide more, since he would want assistance in large, unfamiliar environments.

\subsection{Behavior Towards the Guide} \label{sec: Behavior Towards the Guide}

When exploring parks alone, participants were utilitarian towards the guide. They spoke to it with direct commands (e.g., “Describe what’s in front of me”), called it “guide” or “assistant,” and only interacted with it when they needed assistance. Some participants (Xiran, Aiden, Len, Isaiah) briefly used polite or respectful language with the guide, typically when they first encountered it. Interestingly, they also treated it with respect after experiencing technical challenges, as if trying to coax the guide to behave better. However, these polite exchanges lasted only a few sentences before participants slipped into utilitarian commands.

To our surprise, nearly all participants (n = 12) became friendlier to their guide when using it around others. They used the guide’s name when speaking to it or came up with their own nicknames for it, used gendered pronouns when talking about it, and even encouraged the confederates to interact with their guide. For example:

\begin{quote}
Confederate 2:  What is that robot following you?\\
Len: This robot following me is my guide. It helps me get all around this space. And yeah, he's just my guide.\\
Confederate 2: Does he have a name or anything?\\
\textbf{Len: No, I don't think he has a name. In my head, I call him Jerry, though.}\\
Confederate 1: Hi, Jerry!\\
\textbf{Len: Oh, let's see if he responds!} \textit{[Len releases the button to send the confederate’s greeting to the guide. He speaks to the confederates again.]} \textbf{Might have given him too much info there.}\\
Guide: Hello. How may I assist you in this park?\\
\textbf{Len: There you go! He said hello back! That's awesome.} Alright. So I'm gonna ask him to walk me over to the yellow roof gazebo first. \textit{[To the guide]} \textbf{Guide me to the yellow gazebo, Jerry.}
\end{quote}

Participants were more likely to address the guide by a name when they were in groups, as seen in the quote above. Nine participants (Niya, Nikolai, Sara, Xiran, Len, Siobhan, Isaiah, Jamal, Jiwoo) addressed the guide as “Giddy” in group settings, compared to just two participants (Nikolai, Sara) who used the guide’s name when exploring alone.

While some participants called the guide by its assigned name, six participants gave the guide nicknames in group settings. These nicknames included “Prince” (Teo), “Rufus” (Aiden), “Jerry” (Len), “Gabby” (Jamal), “Robey” (Jamal), and “Diego” (Yadriel). Four nicknames (“Prince,” “Rufus,” “Gabby,” and “Diego”) were for the dog guide, while only two (“Jerry” and “Robey”) were for the robot.

Most participants (n = 10) also began role-playing with the guide based on its appearance when in group settings. They referred to their guide as a “dog” or “robot” instead of just “guide” or “assistant.” Some even changed their language when complimenting or critiquing it, for instance, praising the dog guide as a “good boy” when it gave good descriptions:

\begin{quote}
Confederate 1: Wow, this time you have a guide [that] described the park really well.\\
\textbf{Len: Yeah, I do. I have, it's a good it's…the goodest dog.}\\
Confederate 1: The goodest dog? He's a good boy?\\
Len: Yeah, yeah. \textit{[laughter]}
\end{quote}

Seven participants (Niya, Maritza, Aiden, Len, Siobhan, Li, Jamal) role-played with the guide as a way to rationalize its mistakes in front of others. For example, when Maritza tried to get the guide to lead her and the confederates to a fountain, the guide either misinterpreted her requests or did not respond. She rationalized the guide’s behavior as follows:

\begin{quote}
\textbf{Maritza:} We're not going anywhere. My dog is not take, my dog is not taking me. \textit{[To the guide]} Take me to the Southern Gazebo. \textit{[To the confederates]} \textbf{My dog went to sleep.} \textit{[laughter]}\\
Confederate 1: I mean, do we need to wake him up? Poor guy.\\
\textbf{Maritza:} \textit{[To the guide]} Take us to the Southern Gazebo. \textit{[To the confederates]} \textbf{I forgot to feed her. She is hungry and on strike}… \textit{[To the guide]} \textbf{Are you here, doggie? No, no, doggy.}\\
Confederate 1: Sorry your dog is on strike.
\end{quote}

Two participants (Aiden, Li) role-played with the guide to rationalize their \textit{own} mistakes. When Aiden accidentally released the button to grab the guide while it was navigating, he explained the issue to the confederates as follows:

\begin{quote}
Aiden: \textit{[To the guide]} Walk me to the Northern Gazebo.\\
Guide: Understood. Grab onto me, and I will take you to Northern Gazebo.\\
\textit{[Aiden accidentally releases the button to hold onto the guide, and the guide continues walking to Northern Gazebo on its own.]}\\
Confederate 1: Did your dog just leave without you?\\
\textbf{Aiden: Yeah, I think he kind of did. His harness, his harness slipped out of my hand.}
\end{quote}

Overall, role-playing behavior was most prominent when the guide was in dog form. Some participants (Anika, Sara, Maritza, Aiden, Jamal, Jiwoo) referred to the guide as a robot when in groups. However, they did not encourage the confederates to interact with it, call it by name, or make robot-related compliments or critiques. Participants typically dismissed the robot’s mistakes or claimed it was not listening, rather than role-playing to rationalize errors. While the above behavioral patterns were consistent across participants, potential confounding factors behind these behaviors are noted further in the following section.

%% file: request_type.tex
\begin{figure}
\includegraphics[width=250pt]{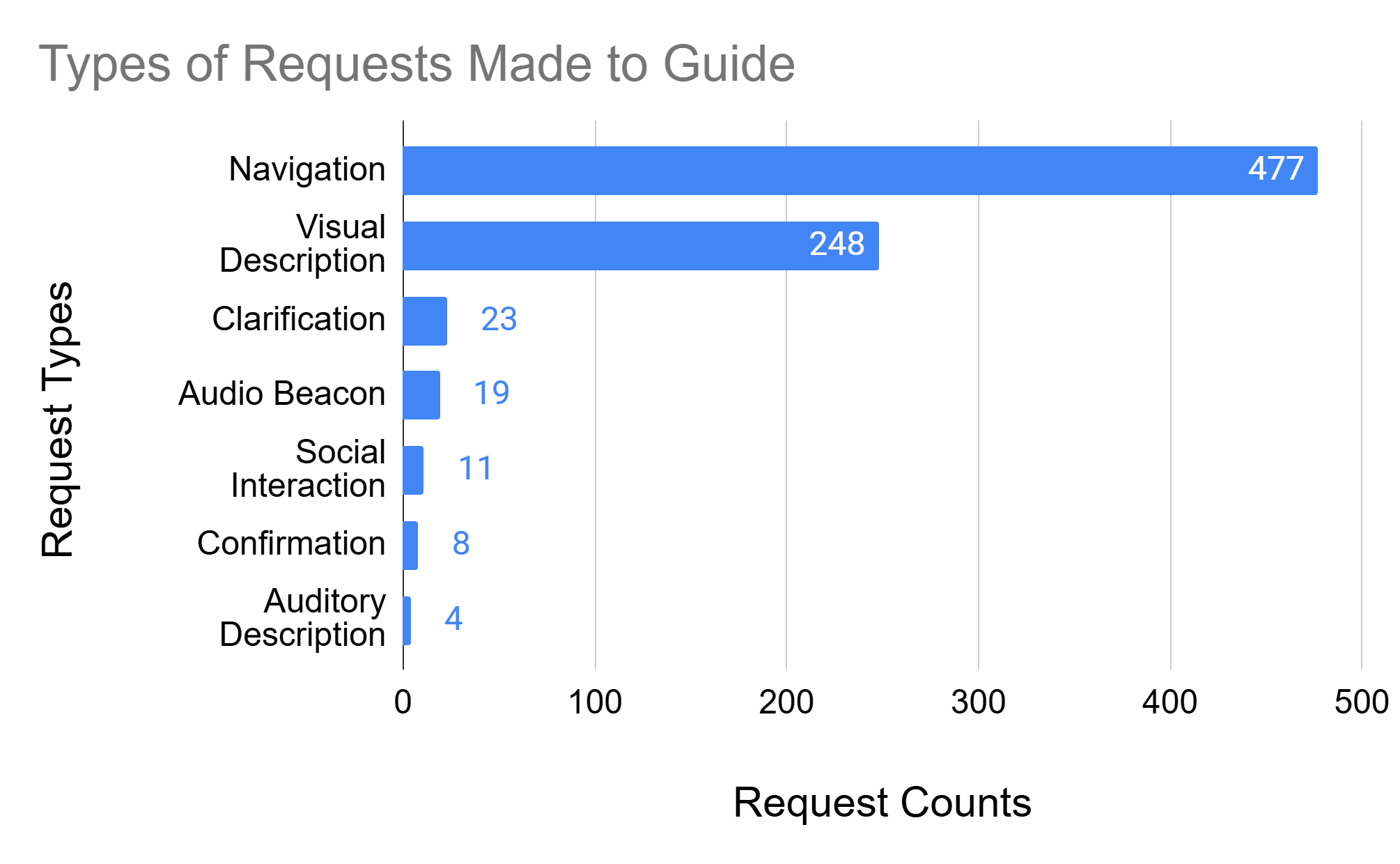}
\caption{\label{fig: request_type} Types of requests made to the guide. Participants made a total of 799 individual requests within 467 user queries, as many queries encompassed multiple request types. For instance, the query “Describe where that sound is coming from” includes both a visual description and an auditory description.
}
\Description{A bar chart depicting the types of requests users made to the guide.

The bar chart is titled "Types of User Requests Made to the Guide." It shows seven categories on the vertical axis labeled "Request Types", and runs from 0 to 500 on the horizontal axis labeled "Request Counts." In order from top to bottom, and greatest to least, the bar chart shows bars for: "Navigation" (477 requests), "Visual Description" (248 requests), "Clarification" (23 requests), "Audio Beacon" (19 requests), "Social Interaction" (10 requests), "Confirmation" (8 requests), and "Auditory Description" (4 requests).
}
\end{figure}

%% file: request_tone.tex
\begin{figure}
\includegraphics[width=250pt]{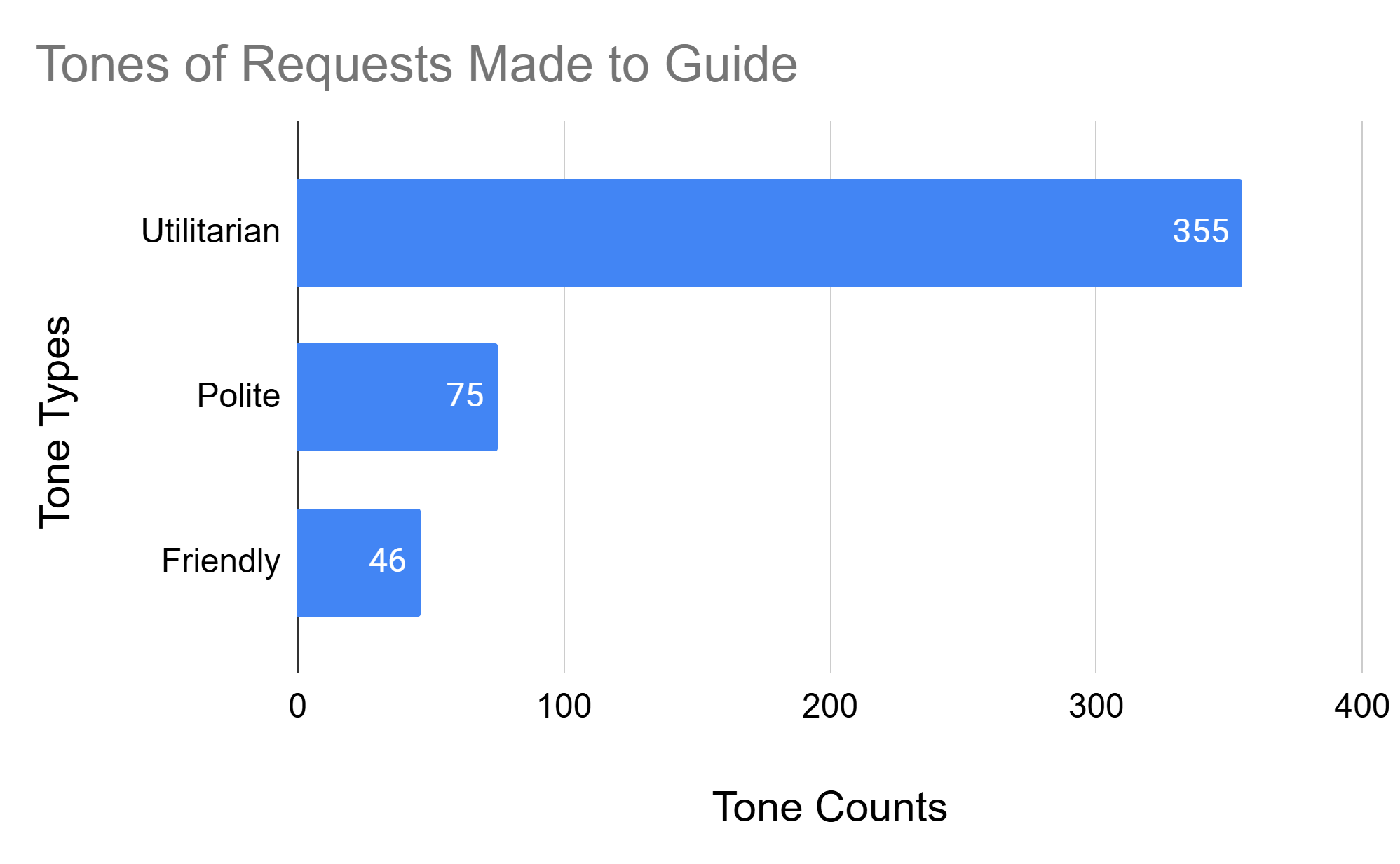}
\caption{\label{fig: request_tone} Tones of requests made to the guide. Participants made 476 requests total, one for each query to the guide, as each query counted as a separate utterance with its own tone.}
\Description{A bar chart depicting the different tones of requests users made to the guide.

The bar chart is titled "Tones of Requests Made to the Guide." It shows three categories on the vertical axis labeled "Tone Types", and runs from 0 to 400 on the horizontal axis labeled "Tone Counts." In order from top to bottom, and greatest to least, the bar chart shows bars for: "Utilitarian" (355 requests), "Polite" (75 requests), and "Friendly" (46 requests).
}
\end{figure}

%% file: likert_responses.tex
\begin{figure*}[t]
\centering
\includegraphics[width=350pt]{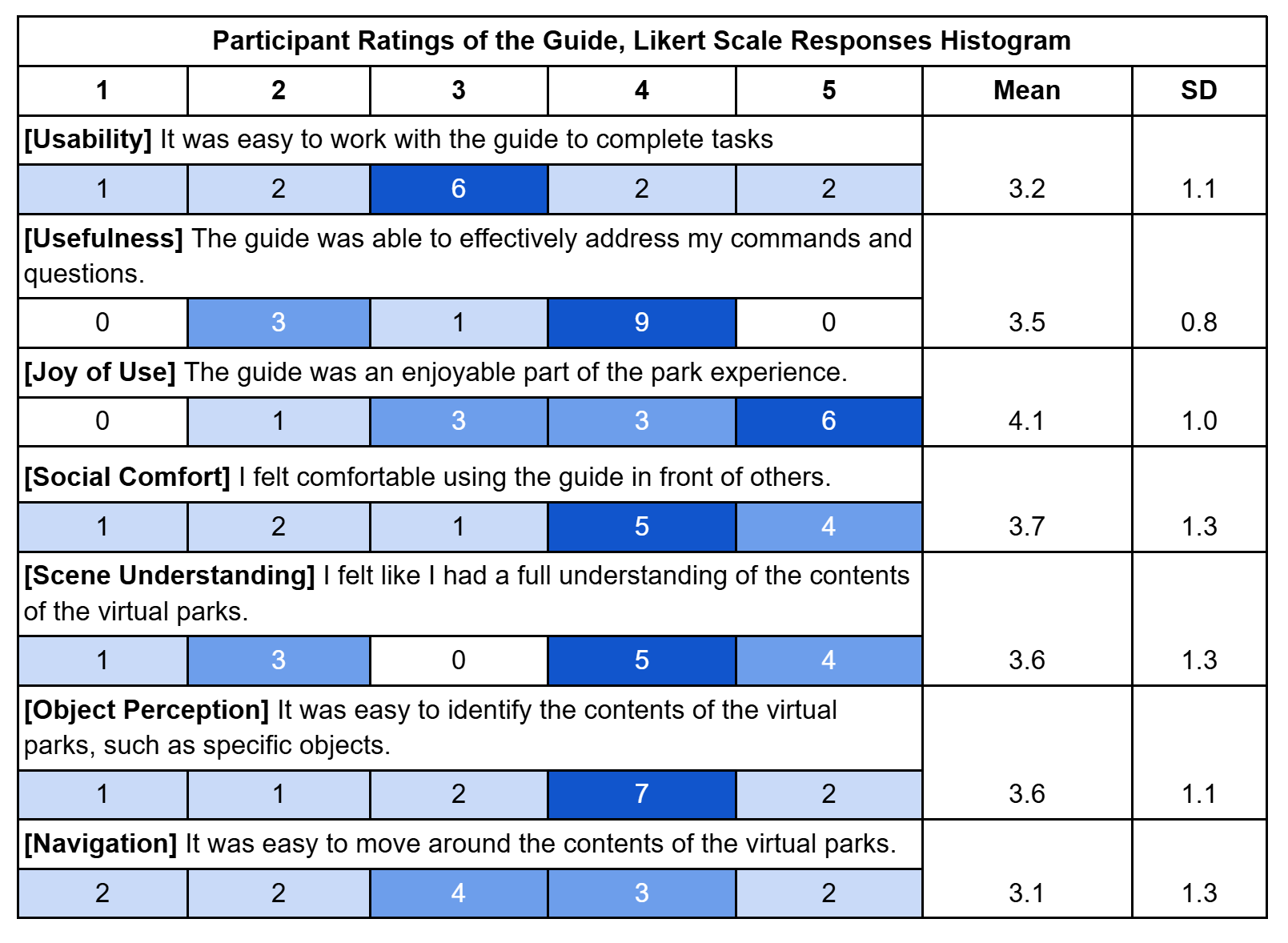}
\caption{\label{fig: Likert Guide Ratings} Likert scale responses from participants rating the guide, collected after participants experienced both parks and guide personas. Scores were given on a scale of 1 to 5, where 1 = strongly disagree, 2 = disagree, 3 = neither agree nor disagree, 4 = agree, and 5 = strongly agree. Note: In the statement on Social Comfort, “others” refers to the confederates participants met during the study. We only include scores from thirteen participants who used the guide.
}
\Description{A histogram containing the Likert-scale response data from participants rating the guide on various qualities. The scale runs from 1 to 5, with numbers in columns below depicting how many responses from participants fell under each number on the scale for a given statement and factor. They are colored in various shades of blue, with the darker colors representing more responses falling under a particular number on the scale and the lighter colors representing fewer.

For the statement: [Usability] It was easy to work with the guide to complete tasks.
The histogram reports 1 user responded 1, 2 users responded 2, 6 users responded 3, 2 users responded 4, and 2 users responded 5. The mean is 3.2, standard deviation 1.1.

For the statement: [Usefulness] The guide was able to effectively address my commands and questions.
The histogram reports 0 users responded 1, 3 users responded 1, 1 user responded 3, 9 users responded 4, and 0 users responded 5. The mean is 3.5, standard deviation 0.8. 

For the statement: [Joy of Use] The guide was an enjoyable part of the park experience.
The histogram reports 0 users responded 1, 1 user responded 2, 3 users responded 3, 3 users responded 4, and 6 users responded 5. The mean is 4.1, standard deviation 1.0. 

For the statement: [Social Comfort] I felt comfortable using the guide in front of others.
The histogram reports 1 user responded 1, 2 users responded 2, 1 user responded 3, 5 users responded 4, and 4 users responded 5. The mean is 3.7, standard deviation 1.3. 

For the statement: [Scene Understanding] I felt like I had a full understanding of the contents of the virtual parks.
The histogram reports 1 user responded 1, 3 users responded 2, 0 users responded 3, 5 users responded 4, and 4 users responded 5. The mean is 3.6, standard deviation 1.1. 

For the statement: [Object Perception] It was easy to identify the contents of the virtual parks, such as specific objects.
The histogram reports 1 user responded 1, 1 user responded 2, 2 users responded 3, 7 users responded 4, and 2 users responded 5. The mean is 3.6, standard deviation 1.1. 

For the statement: [Navigation] It was easy to move around the contents of the virtual parks.
The histogram reports 2 users responded 1, 2 users responded 2, 4 users responded 3, 3 users responded 4, and 2 users responded 5. The mean is 3.1, standard deviation 1.3. 

}
\end{figure*}

%% file: 6Discussion.tex
\section{Discussion}

To our knowledge, this paper presents the first exploration of BLV people’s experience with an AI guide in VR. This work successfully addresses our two overarching research questions by providing key insights into both the utilization (RQ1) and behaviors (RQ2) of BLV users towards an AI guide. Regarding RQ1, participants successfully used the guide to explore and gather reliable information about complex virtual environments. Some were even able to lead sighted groups on tours within those environments, effectively balancing navigation with social interaction. Regarding RQ2, we found that participants altered their treatment of the guide according to social contexts. While alone, they treated the guide as a tool to address their visual needs, keeping interactions brief and direct. However, in social situations, participants became more friendly and role-played with the guide according to its persona.

It should be noted that the behaviors we observed likely reflect a complex interplay between the guide’s design, social presence, and technical or environmental factors (e.g., system latency encouraging name-calling or “attention-grabbing” behaviors from users). Such behaviors offer a compelling foundation for future controlled experiments to hypothesize about and isolate the specific variables behind users’ actions towards an AI guide. Ultimately, these observations and our other insights are critical for developing future AI guides that are not only functionally effective but socially adaptable, enhancing accessibility and user experience for BLV individuals in diverse virtual environments.

\subsection{Comparing Human and AI Guides} \label{sec: Comparing Human and AI Guides}

We compared our results to our previous work on human guides in VR \cite{collins2023guide}, noting interesting differences between experiences with human and AI guides. While participants in both studies formed emotional connections with their guides, our AI-guided participants also displayed novel, AI-specific uses, suggesting a need for new features exclusive to AI guides.

We discuss our comparison with some caveats. For example, our human guide study did not include social situations where participants used the guide around confederates. Further, while people are typically experienced with human conversation, they are far less experienced interacting with an AI. Lastly, although participants in both studies were first-time guide users, using the human guide was closer to real-world conversation, whereas the AI guide required additional training for features like a button interface and audio beacons. These factors may have affected how users adapted to the guides and employed them in the studies.

Regardless of differences, both studies demonstrate that forming emotional connections with guides can be an enjoyable part of guidance. Users of the human guide often invited their guides to perform activities in VR with them, while users of the AI guide role-played with the guide to incorporate it playfully into their experience. These behaviors reflect prior research on sighted guides \cite{guerreiro2019cabot, ball2022runners}, assistive robots \cite{bonani2018robots}, and existing work that has shown people tend to form relationships with AI \cite{zimmerman2023humanai, gur2025algorithm}. Participants leveraged features like the inviting avatars of the virtual guides to develop a degree of closeness, though a lack of other desired features—like the guide dog reacting to being petted (Sara)—limited this connection.

We also observed differences in how participants used the AI guide compared to a human one. Beyond its primary function as a navigational assistant, the AI guide was used as a \textbf{general knowledge assistant}. Participants frequently asked simple, factual questions like, "What is a gazebo?" This contrasts with the human guide, which was used as a \textbf{technical support specialist} and asked more complex questions about the VR system and technical issues. These different uses are revealing: participants saw the AI as a reliable source for simple information, a digital tool for rote inquiries that they would not want to burden human guides with.

Finally, we observed key differences in how participants viewed the roles of human and AI guides during error prevention and recovery. Participants in the human guide study expected their guide to be \textbf{proactive}, offering preemptive warnings to avoid virtual hazards. In contrast, participants treated our AI guide as a \textbf{reactive failsafe}. They did not ask it to prevent mistakes but instead used it to recover from them, such as by asking it to teleport them to a safe location after becoming lost. This suggests that users hold AI guides to a different standard, expecting them to be tools for recovery rather than prevention.

Based on our findings, we propose the following design recommendations for virtual guides:

\begin{itemize}
\item \textbf{Foster Emotional Connections}: Future AI guides could incorporate interactive capabilities (see section \ref{sec: Advantages of Embodied, Personable Accessibility Assistants} to respond appropriately to role-playing.
\item \textbf{Encourage Advanced Usage}: Future AI guides should incorporate features that explicitly demonstrate their capacity for complex tasks. This could include having the guide proactively offer visual details or check on users who appear to be struggling.
\item \textbf{Align with User Roles}: Future AI guides should offer explicit failsafe mechanisms. Rather than having users give voice commands requesting the guide to teleport them to certain locations when they get stuck or lost, users could perform a simple button press that instantly returns them to a safe, known location.
\end {itemize}

\subsection{Advantages of Embodied, Personable Accessibility Assistants} \label{sec: Advantages of Embodied, Personable Accessibility Assistants}

Prior approaches to VR accessibility have focused on informative tools activated through precise controls, similar to screen readers. For example, Microsoft’s Scene Weaver \cite{balasubramanian2023sceneweaver} conveys information about avatars and interactable objects through customizable haptic and audio cues, or manually-triggered alt text. In contrast, many commercial VR applications employ virtual assistants imbued with personality \cite{nathie2019presence, oculus2019first, moss2018review, ven2021vr, cosmonious2022walkthrough}. These assistants are often comical or endearing companions guiding users through tutorials, enhancing engagement and fostering emotional connections between user and assistant \cite{nathie2019presence, cosmonious2022walkthrough, lakereddit, fink2022review, graf2022emotional, clavel2013artificial, sutskova2023cognitive, freeman2025comforting}.

Despite their success in mainstream VR, such personality-infused assistants have not been widely applied in accessibility contexts. However, emerging research indicates potential benefits \cite{phutane2023speaking}. Phutane et al. explored personality-driven conversational assistants for BLV users, reporting positive responses to their interactivity \cite{phutane2023speaking}. Building on this, our AI guide is an embodied agent with personality, assisting users through conversation rather than complex controls. Our observations of participant-guide interactions revealed several advantages of this approach. Namely, the guide leveraged users’ natural desire to role-play with assistive technology, which allowed them to rationalize AI mistakes, reduce embarrassment, and encourage positive social interactions. For example, when the guide appeared in its dog form, confederates pet it, mirroring the social benefits reported by guide dog handlers in the physical world \cite{hwang2024towards}.

Future AI guide designs should support role-playing behaviors beyond mere speech modifications to enhance the user experience. For instance, an AI guide in dog form could play fetch, while a robot form could interact with other technologies. Future studies should also explore BLV users' reactions to various guide embodiments, as we found the guide's appearance significantly influenced the extent of user role-playing. Future research could explore how users respond to guides resembling characters from popular media or traditional assistive devices, like white canes. Such research could also improve how users learn to work with guides. For instance, using a well-liked dog form in tutorials may foster patience during the learning process.

\subsection{Improving Interactions with AI Guides} \label{sec: Improving Interactions with AI Guides}

Although participants had been told they could frame their questions to the guide in any way they wanted, participants tended to re-use simplistic orders (e.g., “Take me to X”, “Describe what’s in front of me”). For example, although several participants disliked the guide’s inconsistent manner of describing their surroundings, none of them tried prompting the guide to give descriptions in specific ways.

This reflects current work on how non-expert users work with modern-day LLMs \cite{zamfirescu2023johnny, brachman2024knowledge, ha2024clochat}. Non-experts typically over-generalize their instructions, and do not provide explicit guidance on how they want the LLM to respond \cite{zamfirescu2023johnny}, leading to less-than-ideal output. 

However, the design of the AI guide's personas may have misled participants as well. For example, when Len told the guide, "Let's go smell the flowers," he likely expected the talkative guide to remember the flower patch it had just mentioned. The guide's verbosity may have implied a conversational understanding and memory that it did not possess. To better align user expectations with the guide's abilities, future versions could include having the guide's appearance reflect its capabilities, such as a small puppy guiding the user rather than a full-grown dog, to indicate inexperience and encourage users to “train” it with more specific commands. The guide could also encourage users to be more explicit with their queries, proactively teaching them how to use it most effectively.

\subsection{Limitations and Future Work} \label{sec: Limitations and Future Work}

Our study was conducted during a period of rapid evolution in generative AI. We started our prototyping and user studies in mid-2023, shortly after the release of GPT-4. While the landscape of AI tools has changed drastically since then—with new APIs offering superior performance and the public’s familiarity with LLMs having grown exponentially—our work demonstrates that prototyping with contemporary tools is necessary to generate lasting insights into human-AI interaction.

Despite its value, our guide prototype faced several architectural limitations due to the systems available at the time. One such drawback was the guide's delayed response time. This delay is an inherent challenge when utilizing external, cloud-based LLMs like GPT-4, as the processing pipeline involves (1) capturing and transmitting user audio and environmental snapshots, (2) processing the query with GPT-4, and (3) generating and streaming the audio response back to the VR headset. Our delays were further exacerbated because, at the time of our study, we lacked access to real-time response streaming. We were forced to wait for the entire response to be generated before transmitting the complete audio file, a constraint that would be largely mitigated by modern streamable Text-to-Speech APIs (e.g., PlayHT \cite{playhtapi}).

Another architectural limitation was the guide's low accuracy (63.2\% across all queries). We attributed many inaccurate responses to technical issues, including the misinterpretation of participant accents, incomplete user queries, or unresponsive APIs. Modern LLMs and Speech-to-Text models offer greater robustness to accent differences. Furthermore, modern APIs often allow for including the context of past queries in LLM calls, which would significantly improve the guide's ability to infer appropriate responses to vague or incomplete prompts. Introducing these features into a modern AI guide architecture would likely enhance its accuracy. Future work should examine which types of guide failures still occur despite these modern advancements and how they impact the user experience, as seen in prior work \cite{goetsu2020different}.

Beyond technical limitations, our study found that participants primarily used a naive approach to prompting the guide. At the time of our study, the general public was becoming aware of AI tools like ChatGPT, but usage was still not widespread. Users today are likely more experienced with prompt engineering and adapting their language to an AI’s perceived capabilities. Future work should include a longitudinal study to examine how users build expertise in crafting sophisticated queries, such as "Describe what’s in front of me using cardinal directions," and how this expertise impacts their perception of the guide's effectiveness. To support this evolving user base, we propose the following recommendations for future AI guides that can help new and experienced users craft effective queries:

\begin{itemize}
\item \textbf{Encourage Advanced Usage}: Guides should actively assist users in prompting them, offering examples of better prompts or teaching users prompt-engineering best practices to help users move beyond basic queries faster.
\item \textbf{Align Personas with Capabilities}: The guide's persona should visually and verbally reflect its capabilities. For instance, a robot persona might encourage blunt, direct language, while a human-like guide could inspire complex commands from users seeking a conversational expert.
\end{itemize}

As AI continues to rapidly evolve and render older implementations obsolete, it remains crucial and valuable to prototype with these developing tools. Findings generated about user behavior and interaction patterns with AI often possess a longevity that the underlying technology lacks. Research on human-AI interaction, including studies on simple voice assistants like Amazon’s Alexa, established enduring trends like the tendency to anthropomorphize AI or rely on it for emotional support \cite{lopatovska2018, gao2018}. Similarly, we expect the user behavior findings from our study—such as the tendency to roleplay with AI assistants or alter communication styles in solitary versus social settings—to carry forward. These findings can inform the design of future, more powerful AI tools, regardless of the architectural advantages that older prototypes might miss.

%% file: 7Conclusion.tex
\section{Conclusion}

We developed an AI guide prototype based on prior work and studied it in social VR environments with 16 BLV people. Participants found the guide was effective for navigation, interpretation, and social interactions, though its appearance and speech characteristics influenced how well participants worked with it. Participants treated the guide differently in solo and social contexts, with some role-playing with the guide and treating it more companionably when other people were around. Participants also revealed behaviors unique to utilizing an AI guide, such as utilizing it for rote memorization and learning terms for objects in their environments. Our findings offer insights for improving AI guides and highlight how embodiment influences their use. Ultimately, this work lays the foundation for designing AI guides that are not just effective tools, but adapt to the complexities of BLV users’ needs and expectations in VR.